\documentclass[a4paper, 12pt]{article}
\usepackage{epsfig,amssymb,euscript,xspace, color}
\usepackage{amsmath,jheppub,mathtools,empheq,amsthm}

\usepackage[T1]{fontenc} 
\usepackage{tikz,caption,subcaption,marvosym} 
\usetikzlibrary{decorations.markings,arrows,snakes}

\def\bea{\begin{eqnarray}}
\def\eea{\end{eqnarray}}
\def\be{\begin{equation}}
\def\ee{\end{equation}}

\definecolor{lightblue}{rgb}{.1,.4,.5}
\definecolor{brown1}{rgb}{.64,.43,.138}



\title{Constructing Carrollian CFTs}
\author{Nishant Gupta and  Nemani V. Suryanarayana}
\emailAdd{nishantg, nemani@imsc.res.in}
\affiliation{Institute of Mathematical Sciences, \\ Taramani, Chennai 600 113, India \\ \\ Homi Bhabha National Institute,  \\ Anushakti Nagar, Mumbai 400085, India}

\abstract{We construct classical theories for scalar fields in arbitrary Carroll spacetimes that are invariant under  Carrollian diffeomorphisms and Weyl transformations. When the local symmetries are gauge fixed these theories become Carrollian conformal field theories. We show that generically there are at least two types of such theories: one in which only time derivatives of the fields appear and the other in which both space and time derivatives appear. A classification of such scalar field theories in three (and higher) dimensions up to two derivative order is provided. We show that only a special case of our theories arises in the ultra-relativistic limit of a covariant parent theory.}

\begin{document}
\maketitle

\section{Introduction}
\label{sec1}

The Poincare algebra admits two interesting limiting algebras obtained by Inonu-Wigner contraction where one takes the speed of light $c \in [0,\infty)$ to either zero or infinity. When $c\rightarrow \infty$ one obtains the well-known Galilean algebra, and when $c \rightarrow 0$ one ends up with the so-called Carrollian algebra \cite{LevyLeblond, SenGupta}. The former case has played a very important role as it is relevant for a host of physical systems in which typical velocities involved are very small compared to the speed of light, such as Newtonian mechanics, many condensed matter systems etc. The Carroll limit is relevant if velocities involved are all very close to $c$ and it remained unexplored for a long time until the last decade. 

An important realisation of the Poincare algebra arises from the isometries of the flat Minkowski spacetime with the standard line element:
\bea
\label{flatmink}
ds^2 = - c^2 \, dt^2 + d{\bf x} \cdot d{\bf x} \nonumber
\eea
which is an example of a pseudo-Riemannian manifold, with non-degenerate Lorentzian metric $g_{\mu\nu} = \eta_{\mu\nu}$. If one takes either of the limits $c \rightarrow \infty$ (better done on $g^{\mu\nu}$) or $c\rightarrow 0$ (done on $g_{\mu\nu}$) one obtains a manifold with a degenerate metric tensor. The geometries that realise Galilean or Carrollian algebras are therefore not Riemannian manifolds but in fact belong to a more general class of geometrical objects called Newton-Cartan manifolds (for $c\rightarrow \infty$ case) and Carroll manifolds (for $c\rightarrow 0$ case) respectively (see, for instance, \cite{Duval:2014uoa}).

The conformal algebra contains the Poincare algebra as a subalgebra, and also plays an important role in physics as it is relevant in many contexts, such as: symmetry algebra of theories at the fixed points of RG flows, AdS/CFT etc. Hence the Galilei and Carroll limits of the Conformal algebra have also been studied in the past. 

Just as the conformal algebra can be represented by conformal Killing vectors (CKV) of (conformally) flat spacetimes, such as ${\mathbb R} \times S^d$ with line element 
\bea
ds^2 = -c^2 \, dt^2 + d\Omega_d^2 \nonumber
\eea
or ${\mathbb R}^{1,d}$, the Galilean and Carrollian conformal algebras can be thought of as appropriately defined conformal algebras of (conformally) flat Galilei and Carroll spacetimes. The conformal algebra of (any spacetime conformal to) Minkowski spacetime $M_{d+1}$ is finite dimensional for $d\ge2$ and therefore obtaining the corresponding Galilean or Carrollian conformal algebras via Inonu-Wigner contraction will necessarily result in finite dimensional algebras for $d\ge 2$. However, if one defines these algebras directly as conformal algebras of the corresponding Galilei and Carroll spacetimes one may get bigger algebras than those obtained by the contraction procedure. In fact there is a more general notion of conformal symmetry even in the flat Carroll spacetime $C_{d+1}$, parametrised by $z$, the analog of the dynamical exponent of Galilean conformal transformations, which in turn is given by a non-negative integer $k$ by $z=2/k$ in \cite{Duval:2014lpa}. One obtains only the $z=1$ ($k=2$) case via the contraction procedure. 

Also as was shown in \cite{Duval:2014uva} the Carrollian conformal algebra $\mathfrak{cca}^{(z)}_{d+1}$ for $d=2$ and $k=2$ is the $\mathfrak{bms}_4$ algebra \cite{Bondi:1962px, Sachs:1962zza}, which is infinite dimensional. This is because the asymptotic null boundary ${\mathcal I}^\pm$ of flat spacetime with metric
\bea
ds^2 = -0 \times  dt^2 + d\Omega_d^2 \nonumber
\eea
is a Carroll manifold. It was well-known that the asymptotic symmetry algebra of 4-dimensional gravity (without cosmological constant) is the famous $\mathfrak{bms}_4$ algebra \cite{Bondi:1962px, Sachs:1962zza}. 

So if one wants to describe some $(d+2)$-dimensional gravitational theory with asymptotically flat boundary conditions holographically then the holograms should be field theories on ${\mathcal I}^\pm$ with $\mathfrak{cca}^{(z=1)}_{d+1}$ as their global symmetries \cite{Barnich:2010eb, Barnich:2009se}. This has led to a lot of work seeking field theories on null-manifolds which are Carrollian CFTs (see for example \cite{Duval:2014uoa, Duval:2014uva, Duval:2014lpa, Ciambelli:2019lap, Ciambelli:2018wre, Ciambelli:2018xat, Bagchi:2019clu, Bagchi:2019xfx, Bagchi:2016bcd, Ciambelli:2018ojf}). The procedure followed there is typically to start with a CFT on $M_{d+1}$ and take an ultra-relativistic limit. This has been a fruitful exercise and has resulted in quite a lot of interesting examples with $\mathfrak{cca}^{(z=1)}_{d+1}$ symmetries. 

It is not inconceivable that constructing field theories directly on a Carroll manifold ${\cal C}_{d+1}$ may give rise to more general class of theories than those obtained by taking ultra-relativistic limit of known CFTs on the parent pseudo-Riemannian manifold ${\cal M}_{d+1}$. The main aim of this paper is to demonstrate that this expectation is indeed realised. We will construct classical scalar field theories on generic 3-dimensional Carroll manifolds that are invariant under both diffeomorphisms and Weyl transformations of the Carroll manifold. These theories can be put on any given Carroll manifold -- of which the interesting examples include null boundaries of asymptotically flat spacetimes, horizons of black holes, boundaries of causal developments, etc. Then the residual symmetries should, by construction, make the resultant theory have $\mathfrak{cca}_3^{(z)}$ as the symmetry algebra for generic values of $z$. We will mainly concentrate on $d=2$ for most part (relegating the higher dimensional case to an Appendix).  We obtain a larger class of theories for scalar fields going beyond what one obtains by the method of taking ultra-relativistic limit. 

\subsection*{Summary of results:}

On a generic 3-dimensional Carroll manifold we have constructed diffeomorphism and Weyl covariant equations of motion as well as invariant actions (when they exist) of a scalar field $\Phi(t, {\bf x})$ up to second order derivatives in both time ($t$) and space (${\bf x}$) coordinates, for general values of $z$ and the conformal dimension $\delta$ of $\Phi(t,{\bf x})$. Our results include:  
\begin{itemize}
\item Two classes of diffeomorphic and Weyl covariant equations of motion, one with  two time-derivatives of the field, and the other with (up to) two space-derivatives. These exist for general values of $z$ and $\delta$.
\item For the special value $z=1$ there is one equation of motion with at least five real parameters.
\item For $z=2$ there are two classes of equations of motion: one with second order time-derivatives and the other with first order time- and second order space-derivatives.
\item The invariant actions exist only when $\delta$ is restricted to either $\delta =\frac{z}{2}$ or $\delta =1-\frac{z}{2}$. 
\item Having a stable monomial potential ($\Phi^{2n}$) further restricts the values of $z$ to be determined in terms of the degree of the monomial. 
\item Gauge fixing the local symmetries in an appropriate way leads to field theories with Carrollian conformal algebra $\mathfrak{cca}_3^{(z)}$ worth of symmetries that have an additional fluctuating two-component field along with the scalar field.
\end{itemize}
The rest of the paper is organised as follows. The Section \ref{sec2} contains a review of how one constructs a diffeomorphic and Weyl invariant theory of a scalar field in the background of a generic (pseudo) Riemann manifold. In Section \ref{sec3} we turn to repeating the steps of Section \ref{sec2} to the case of scalars in a Carroll manifold. There we present the details of our construction of equations of motion as well as actions. In Section \ref{sec4} we show how a subset of results of Section \ref{sec3} can be recovered starting from the conformally coupled scalar with the background metric in Randers-Papapetrou form. In Section \ref{sec5} we gauge fix our theories to recover and generalise the classical scalar field theories with $\mathfrak{cca}_3^{(z)}$ algebra worth of symmetries. We conclude with some remarks and open questions in Section \ref{sec6}. The Appendix \ref{appendix} contains some details of the case of arbitrary dimensions.
\section{Conformally coupled scalar field - revisited}
\label{sec2}
In this Section we (re) construct the well-known conformally coupled scalar field equation of motion and its action in detail. This will provide us with the method we follow in the later sections. The result we want to re-derive is the diffeomorphic and Weyl invariant classical (free) scalar field theory in $d=2+1$ dimensions in a general background with metric $g_{\mu\nu}$. This has the action:
\bea
\label{csaction}
{\cal L} = -\frac{1}{2} \sqrt{-g} \, \left(g^{\mu\nu} \partial_\mu \phi \, \partial_\nu \phi + \frac{1}{8} R\, \phi^2 \right)
\eea
and the equation of motion
\bea
\label{cseom}
g^{\mu\nu} \nabla_\mu \nabla_\nu \phi - \frac{1}{8} R \, \phi =0 \, .
\eea
This action (\ref{csaction}) is invariant and the equation of motion (\ref{cseom}) is covariant under the Weyl transformations:
\bea
\label{rweyl}
g'_{\mu\nu} (x) = \frac{1}{B^2} g_{\mu\nu} (x), ~~ \phi' (x) = B^{\frac{1}{2}} \phi(x)
\eea
where $B$ is an arbitrary function of the coordinates, and diffeomorphisms:
\bea
\label{rdiff}
g'_{\mu\nu} (x') = \frac{d x^\alpha}{dx'^\mu} \frac{d x^\beta}{dx'^\nu} g_{\alpha\beta} (x), ~~ \phi' (x') = \phi(x)
\eea
where $x \rightarrow x'^\mu(x)$ is a coordinate transformation. Let us rederive this result (see Wald \cite{Wald:1984rg} for instance). For this one starts with scalar (under diffeomorphisms in (\ref{rdiff})) combinations that are linear in $\Phi$ and have (at most) two derivatives, namely, $R \, \Phi$ and $\nabla_\mu \nabla^\mu \Phi$. Their transformation properties under the Weyl transformations $g_{\mu\nu} \rightarrow B^{-2} \, g_{\mu\nu}$ and $\Phi \rightarrow B^\delta \, \Phi$, are
\bea
\label{rweyltrans}
R \, \Phi &\longrightarrow& B^\delta \, \Big[ B^2 \, R +4 \, B\, g^{\mu\nu} \nabla_\mu \nabla_\nu B -6 \, g^{\mu\nu} \nabla_\mu B \, \nabla_\nu B \Big] \, \Phi \, ,\cr
g^{\mu\nu} \nabla_\mu\nabla_\nu \Phi &\longrightarrow& B^\delta \, \Big[B^2 \, g^{\mu\nu} \nabla_\mu\nabla_\nu \Phi + \delta \, \Phi \, B \, g^{\mu\nu} \nabla_\mu\nabla_\nu B \, + \delta \, (\delta-2) \, \Phi \, g^{\mu\nu} \nabla_\mu B \, \nabla_\nu B \cr 
&& ~~~~~~ + (2\delta -1) \, B\, g^{\mu\nu} \nabla_\mu B \, \nabla_\nu \Phi \Big] 
\eea
where $\delta$ is the Weyl weight of the scalar $\Phi$. For a linear combination of $R \, \Phi$ and $\square \Phi$ to be covariant all the inhomogeneous terms in the Weyl transformation of that combination should cancel out. But in no linear combination the term containing $B\, g^{\mu\nu} \nabla_\mu B \, \nabla_\nu \Phi$ on the right hand side of $\square \Phi$ in (\ref{rweyltrans}) gets canceled. So its coefficient $(2\delta-1)$ has to vanish identically, giving us $\delta =\frac{1}{2}$. Then the only linear combination that transforms homogeneously is
\bea
g^{\mu\nu} \nabla_\mu\nabla_\nu \Phi - \frac{1}{8} R \, \Phi &\rightarrow& B^{\frac{5}{2}} \, \Big[ g^{\mu\nu} \nabla_\mu\nabla_\nu \Phi - \frac{1}{8} R \, \Phi  \Big]
\eea
showing that the equation (\ref{cseom}) is the only covariant one. For the construction of an action one notes:
\bea
g^{\mu\nu} \nabla_\mu \Phi \, \nabla_\nu \Phi &\longrightarrow& B \Big[ B^2 \, g^{\mu\nu} \nabla_\mu \Phi \, \nabla_\nu \Phi + \frac{1}{4} \Phi^2 \, g^{\mu\nu} \nabla_\mu B \, \nabla_\nu B + \Phi \, B \, g^{\mu\nu} \nabla_\mu \Phi \, \nabla_\nu B \Big]
\eea
which can be used along with the first line of (\ref{rweyltrans}) to show that
\bea
\sqrt{g} \, \left[ g^{\mu\nu} \nabla_\mu \Phi \, \nabla_\nu \Phi + \frac{1}{8} R \, \Phi^2 \right] \longrightarrow \sqrt{g} \, \left[ g^{\mu\nu} \nabla_\mu \Phi \, \nabla_\nu \Phi + \frac{1}{8} R \, \Phi^2 \right]  + \partial_\mu \left[ \frac{\Phi^2}{2 B} \sqrt{g}  \, g^{\mu\nu} \partial_\nu B \right] \, .
\eea
So the action (\ref{csaction}) is also invariant under (\ref{rweyl}, \ref{rdiff}) and results in the equation of motion (\ref{cseom}). In the next section we use the same procedure to construct analogous equations of motion and actions for scalar fields on 3-dimensional Carroll manifolds.
\section{Scalar field theories on Carroll Spacetimes}
\label{sec3}
Let us first review some essential aspects of Carrollian geometries. We will follow notations and conventions of Ciambelli et al \cite{Ciambelli:2019lap}, \cite{Ciambelli:2018xat} here. A Carroll spacetime is a fibre bundle ${\cal C}_{d+1}$ with a $d$-dimensional base ${\cal S}$ and one-dimensional fibre. We work with local coordinates ${\bf x}$ on the base and $t$ on the fibre. Then the Carroll spacetime is specified by a non-degenerate $d$-dimensional metric $a_{ij}(t, {\bf x})$ on the base ${\cal S}$, the Ehresmann connection 1-form $b_i (t, {\bf x})$ and a scalar $\omega(t, {\bf x})$. Then one defines the Carroll diffeomorphisms as those that keep this structure invariant. In our coordinates they take the form:
\bea
\label{cdiff}
t\rightarrow t'(t, {\bf x}), ~~ {\bf x} \rightarrow {\bf x}'({\bf x}).
\eea
The Jacobian of these Carroll diffeomorphisms $(t, x^i) \rightarrow (t'(t, {\bf x}), x'^i({\bf x}))$  is the matrix
\bea
\left( \begin{array}{cc} J(t,{\bf x}) & J_i (t, {\bf x}) \\ 0 & {J^j}_i \end{array} \right)
\eea
where $J= \frac{\partial t'}{\partial t}$, $J_i = \frac{\partial t'}{\partial x^i}$, and ${J^j}_i = \frac{\partial x'^j}{\partial x^i}$, with its inverse
\bea
\left( \begin{array}{cc} J^{-1} & ~~ - J^{-1} J_k {(J^{-1})^k}_i\\ 0 & ~~ {{(J^{-1})}^j}_i \end{array} \right)
\eea
where ${(J^{-1})^i}_j$ is the inverse of the matrix ${J^i}_j$. Under these transformations the geometrical data $(a_{ij}, b_i, \omega)$ of the Carroll spacetime transforms as:
\bea
\label{cdifftrans}
a'_{ij} (t', {\bf x}') &=& a_{kl} (t, {\bf x}) \, {{(J^{-1})}^k}_i {{(J^{-1})}^l}_j, 
~~~ \omega' (t', {\bf x}')  = J^{-1}  \, \omega (t, {\bf x}) \, \, \cr \cr
b'_k (t', {\bf x}') &=& \Big(b_i (t, {\bf x}) + J^{-1}  \, J_i \, \omega (t, {\bf x}) \Big) \, {(J^{-1})^i}_k 
\eea
along with $\partial'_t = J^{-1} \partial_t$, $\partial'_j= {(J^{-1})^i}_j (\partial_i - J^{-1} \, J_i \, \partial_t)$. Now one can list the objects that are covariant under the Carroll diffeomorphisms. At the first derivative order one has 
\bea
\phi_i=\frac{1}{\omega}(\partial_i \omega+\partial_t b_i), ~~ \hat{\gamma}^i_j=\frac{1}{2\omega}a^{ik}\partial_t a_{jk}, ~~ \theta = \frac{1}{\omega} \partial_t \ln \sqrt{a} = {\hat \gamma^i}_i, ~~ f_{ij} =2\, (\partial_{[i} b_{j]}  + b_{[i} \, \phi_{j]} ) \, .
\eea
Because these are covariant one can raise and lower the indices using $a_{ij}$ and its inverse $a^{ij}$. One also has the following differential operators 
\bea
\hat \partial_t = \frac{1}{\omega} \partial_t, ~~ \hat \partial_i = \partial_i + \frac{b_i}{\omega} \, \partial_t
\eea
that are covariant. Then the Carroll-Christoffel connection 
\bea
\label{ccconnection}
{\hat \gamma^i}_{jk} =\frac{1}{2} a^{il} (\hat \partial_j a_{lk} + \hat \partial_k a_{jl} - \hat \partial_l a_{jk})
\eea
allows one to write down further sets of covariant objects.  This connection transforms under Carroll diffeomorphisms in the same manner as the usual Christoffel connection in Riemannian geometry. We will define the Carroll tensors to transform as:
\bea
\Phi' = \Phi, ~~ V'^i = {J^i}_j \, V^j, ~~ V'_i = V_j {(J^{-1})^j}_i, ~~ etc.
\eea
Under the Carroll connection (\ref{ccconnection}) $a_{ij}$ and $a^{ij}$ are covariantly constant. Another fact is that if one defines $\hat \partial_t := \frac{1}{\omega} \partial_t$ and $\hat \nabla_t a_{ij} := \hat \partial_t a_{ij} - \hat \gamma^k_i a_{kj} - \hat \gamma^k_j a_{ik}$ and $\hat \nabla_t a^{ij} := \hat \partial_t a^{ij} + \hat \gamma_k^i a^{kj} + \hat \gamma_k^j a^{ik}$ then the metric $a_{ij}$ and its inverse are covariantly constants under $\hat \nabla_t$ as well. 

At the second derivative order one can list the following invariant objects:
\bea
\label{cdiffinvs}
\theta^2, ~~ \hat \partial_t \theta = \frac{1}{\omega} \partial_t \theta, ~~ \hat \gamma^i_j \hat \gamma^j_i, ~~ \hat r, ~~ a^{ij} \hat \nabla_i \phi_j, ~~ a^{ij}  \phi_i \phi_j
\eea
where $\hat r$ is the Carroll Ricci scalar defined \cite{Ciambelli:2018xat} as follows:
\bea
{\hat r^i}_{jkl} &=& \hat \partial_k \hat \gamma^i_{lj} - \hat \partial_l \hat \gamma^i_{kj} + \hat \gamma^i_{km} \hat \gamma^m_{lj} - \hat \gamma^i_{lm} \hat \gamma^m_{kj}, ~~ \hat r_{ij} =   {\hat r^k}_{ikj}, ~~ \hat r = a^{ij} \hat r_{ij} \, .
\eea
Next we consider Carroll Weyl transformations. Following \cite{Ciambelli:2019lap} we define this as
\bea
\label{cweyl}
\tilde a_{ij}(t, {\bf x})= (B(t, {\bf x}))^{-2} a_{ij}(t, {\bf x}), ~~ \tilde \omega(t, {\bf x}) = (B(t, {\bf x}))^{-z} \omega(t, {\bf x}), ~~~ \tilde b_i(t, {\bf x}) = (B(t, {\bf x}))^{-z} b_i(t, {\bf x}) \cr
\eea
where $B(t, {\bf x})$ is an arbitrary function and $z$ is a non-zero real number.
We are now ready to emulate the steps of section \ref{sec2} and construct equations of motion that are covariant under the Carroll diffeomorphisms (\ref{cdiff}) and Weyl transformations (\ref{cweyl}).

\subsection{Constructing equations of motion}
For this we first start by listing the transformation properties of our Carroll diffeomorphism invariants (\ref{cdiffinvs}) under (\ref{cweyl}). One finds:
\bea
\label{wtrans1}
\theta &\longrightarrow& B^{z-1} \left[ B \, \theta -2 \, \hat \partial_t B \right]  \, ,\cr
\hat \partial_t \theta &\longrightarrow& B^{2\,z-2} \left[ B^2 \, \hat \partial_t \theta + z \, B \, \theta \, \hat \partial_t B - 2 \, (z-1) \, (\hat \partial_t B)^2 -2 B \, \hat \partial_t \hat \partial_t B \right] \, ,\cr 
\hat \gamma^i_j \hat \gamma^j_i &\longrightarrow& B^{2\,z-2} \left[ B^2 \,\hat \gamma^i_j \hat \gamma^j_i -2 \, B \, \theta \, \hat \partial_t B +2\, (\hat \partial_t B)^2 \right] \, ,\cr
\hat r &\longrightarrow& B^2 \, \hat r + 2 B \, a^{ij} \hat \nabla_i \hat \partial_j B -2 a^{ij} \hat \partial_i B \hat \partial_j B \, , \cr
a^{ij} \hat \nabla_i \phi_j &\longrightarrow& B^2 \, a^{ij} \hat \nabla_i \phi_j - z \, B \, a^{ij} \hat \nabla_i \hat \partial_j B + z \, a^{ij} \hat \partial_i B \hat \partial_j B  \,  , \cr
a^{ij} \phi_i \phi_j &\longrightarrow& B^2 \, a^{ij} \phi_i \phi_j -2z\, B \, \phi^i \hat \partial_i B + z^2 \, a^{ij} \hat \partial_i B \hat \partial_j B \, .
\eea
There are three combinations that transform homogeneously:
\bea
\hat r + \frac{2}{z} a^{ij} \hat \nabla_i \phi_j &\longrightarrow& B^2 \left(\hat r + \frac{2}{z} a^{ij} \hat \nabla_i \phi_j \right)  \, ,\cr
\hat \gamma^i_j \hat \gamma^j_i -\frac{1}{2} \theta^2  &\longrightarrow& B^{2 \, z} (\hat \gamma^i_j \hat \gamma^j_i -\frac{1}{2} \theta^2)  \, ,\cr
f_{ij} f^{ij} &\longrightarrow& B^{4-2 \, z} \, f_{ij} f^{ij} \, .
\eea
Now we are ready to include matter fields. Let us consider a real scalar field $\Phi(t, {\bf x})$ for simplicity. We seek to construct equations of motion which are Carroll Weyl invariant. For this we start by listing Carroll diffeomorphism invariants up to two derivatives (on both sets of  objects $(a_{ij}, b_i, \omega)$ and $\Phi$):
\bea
\label{cdinv2}
&& \Phi, ~~ \theta \, \Phi, ~~ \theta^2 \, \Phi, ~~ \hat \gamma^i_j \hat \gamma^j_i \, \Phi, ~~ \hat \partial_t \theta \, \Phi, ~~ \hat r \, \Phi, ~~ \hat \nabla^i \phi_i \, \Phi, ~~  \phi^i  \phi_i \, \Phi, ~~ f_{ij} f^{ij} \Phi \cr
&& \hat \partial_t \Phi, ~~ \theta \, \hat \partial_t \Phi, ~~  \phi^i \, \partial_i \Phi, ~~~ \hat \partial_t \hat \partial_t \Phi, ~~ \hat \nabla^i \hat \partial_i \Phi
\eea
Defining that the scalar $\Phi$ transforms as $\Phi \rightarrow B^\delta \Phi$ under Weyl transformations we can find how the objects in (\ref{cdinv2}) above transform. We find:
\bea
\label{weyltrans}
\hat \partial_t \Phi &\longrightarrow& B^{z+\delta -1} \, \left[ B \, \hat \partial_t \Phi + \delta \, \Phi \, \hat \partial_t B \right] \, ,\cr
\hat \partial_t \hat \partial_t \Phi &\longrightarrow& B^{\delta +  2\, z -2} \left[ B^2 \, \hat \partial_t \hat \partial_t \Phi + (z+ 2\, \delta) \, B \, \partial_t B \, \partial_t \Phi + \delta \, B \, \Phi \, \hat \partial_t \hat \partial_t B + \delta \, (\delta + z-1) \Phi \, (\hat \partial_t B)^2 \right] \, ,\cr
\phi^i \, \hat \partial_i \Phi &\longrightarrow& B^\delta \left[B^2 \, \phi^i \, \hat \partial_i \Phi + \delta \, B \, \Phi \, \phi^i \, \hat \partial_i B -\delta \, z \, \Phi \, \hat \partial_i B \, \hat \partial^i B - z \, B \, \hat \partial^i \Phi \, \hat \partial_i B \right] \, ,\cr
\hat \nabla^i \hat \partial_i \Phi &\longrightarrow& B^\delta \, \left[ B^2 \, \hat \nabla^i \hat \partial_i \Phi + \delta (\delta-1) \Phi \hat \partial^i B \, \hat \partial_i B + 2 \, \delta \, B \hat \partial^i B \, \hat \partial_i \Phi + \delta \, B \, \hat \nabla^i \hat \partial_i B\right] \, .
\eea
Clearly the simplest Weyl covariant object is at the first derivative order :
\bea
\label{1storder}
\hat \partial_t \Phi + \frac{\delta}{2} \theta \, \Phi  \longrightarrow B^{z+\delta} (\hat \partial_t \Phi + \frac{\delta}{2} \theta \, \Phi) \, .
\eea
This combination was already known in \cite{Ciambelli:2018xat}. At the second derivative order we find two such covariant objects:
\bea
\label{1stkind}
&& \hat \partial_t^2 \Phi + \frac{1}{2} (z + 2 \, \delta) \, \theta \, \hat \partial_t \Phi + \frac{\delta}{4} \left[ (z+\delta) \, \theta^2 +2\, \hat \partial_t \theta \right] \, \Phi \cr
&&  \hskip 1cm \longrightarrow B^{2\,z+ \delta} \left(\hat \partial_t^2 \Phi + \frac{1}{2} (z + 2 \, \delta) \, \theta \, \hat \partial_t \Phi + \frac{\delta}{4} \left[ (z+\delta) \, \theta^2 + 2\, \hat \partial_t \theta \right] \, \Phi \right) \, \\
\label{2ndkind}
&& \hat \nabla^i \hat \partial_i \Phi + \frac{2 \, \delta}{z} \, \phi^i \, \partial_i \Phi - \frac{\delta}{2} \left[ \hat r - \frac{2 \, \delta}{z^2} \, \phi^i \, \phi_i \, \right] \, \Phi \cr
&&  \hskip 1cm \longrightarrow B^{2+ \delta} \left(\hat \nabla^i \hat \partial_i \Phi + \frac{2 \, \delta}{z} \, \phi^i \, \hat \partial_i \Phi - \frac{\delta}{2} \left[ \hat r - \frac{2 \, \delta}{z^2} \, \phi^i \, \phi_i \, \right] \, \Phi \right) \, .
\eea
Therefore we have found two distinct possibilities for the covariant equations of motion: one with weight $2z+\delta$ and the other with weight $2+\delta$. We will refer to these as {\it time-like} case and {\it space-like} case respectively. 

These are not yet the most general covariant combinations. One is free to consider linear combination of (\ref{1stkind}) with $(\hat \gamma^i_j \hat \gamma^j_i -\frac{1}{2} \theta^2) \, \Phi$ as this also has weight $2\,z+\delta$. Also, whenever $\frac{2\,z+\delta}{\delta}$ ($:=N_t -1$) is a non-negative integer  then one can also consider $\Phi^{\frac{2\,z+\delta}{\delta}}$ along with (\ref{1stkind}). Such a term is expected to contribute to the equation of motion when there is a monomial potential of the type $\Phi^{N_t}$ in the action for $\Phi$. If such a potential is included, then $\delta = 2z/(N_t-2)$ and if we further demand that $\delta \ge 0$ we need $N_t \ge 3$.

Similarly one can consider a linear combination of (\ref{2ndkind}) with $(\hat r + \frac{2}{z} a^{ij} \hat \nabla_i \phi_j) \, \Phi$ and/or with $\Phi^{\frac{2+\delta}{\delta}}$ whenever $\frac{2+\delta}{\delta}$ ($:=N_s-1$) is a non-negative integer - as these quantities also have Weyl weight $2+\delta$. Again this implies $\delta = 2/(N_s-2)$, and for $\delta \ge 0$ we need $N_s \ge 3$. If we further demand that the monomial potentials are stable then we will have to restrict $N_t$ and $N_s$ to be even integers: $N_t = 2 \, n_t$ and $N_s = 2 \, n_s$ for non-negative integers $n_t$ and $n_s$.

There are two special values of $z$, namely, $z=1$ and $z=2$, that have to be treated more carefully because in these cases we can consider more general covariant combinations.
\begin{itemize}
\item $z=1$:  We can consider constant linear combinations of the two quantities (\ref{1stkind}) and (\ref{2ndkind}) -- for in this case the Weyl weights of both of these quantities become equal to $2+\delta$. Furthermore precisely in this case the Weyl weight of  $f_{ij} f^{ij} \, \Phi$ also becomes $2+\delta$ and so comes into play. 
\item $z=2$: We can consider linear combinations of the first order time-derivative object (\ref{1storder}) with (\ref{2ndkind}).
\end{itemize}
We can summarise our results so far for various types of Carroll diffeomorphic and Weyl invariant equations of motion for a scalar field $\Phi$ as follows:
\begin{itemize}
\item For $z=1$ we can take the covariant equation to be:
\bea
\label{z1eqns}
&&~~\, \kappa_0 \left[\hat \partial_t^2 \Phi + \frac{1}{2} (1 + 2 \, \delta) \, \theta \, \hat \partial_t \Phi + \frac{\delta}{4} \left[ (1+\delta) \, \theta^2 + 2 \, \hat \partial_t \theta \right] \, \Phi \right]   \cr
&& + \,\kappa_1 \, \left[ \hat \nabla^i \hat \partial_i \Phi + 2 \, \delta\,  \phi^i \, \hat \partial_i \Phi - \frac{\delta}{2} \left[ \hat r - 2 \, \delta \, \phi^i  \phi_i \, \right] \, \Phi \right]  \cr
&& + \left[ \sigma_0 \, \left(\hat \gamma^i_j \hat \gamma^j_i -\frac{1}{2} \theta^2 \right) + \sigma_1 \, \left(\hat r + 2 \, a^{ij} \hat \nabla_i \phi_j \right) + \sigma_2 \, f_{ij} f^{ij} \right] \, \Phi + \lambda \, \Phi^{\frac{2+\delta}{\delta}}= 0 \, .
\eea
This equation has five independent real parameters -- since $(\kappa_0, \kappa_1, \sigma_0, \sigma_1, \sigma_2, \lambda)$ are equivalent to $\alpha \, (\kappa_0, \kappa_1, \sigma_0, \sigma_1, \sigma_2, \lambda)$ for any non-zero real $\alpha$. Note also that there is no reason to fix $\delta$ at this stage beyond the assumption that $\frac{2+\delta}{\delta}$ is (preferably an odd) positive integer, whenever $\lambda \ne 0$.

\item For $z=2$ we can take the covariant equation to be:
\bea
\label{z2eqns1}
\kappa_3 \, \left[ \hat \partial_t \Phi + \frac{\delta}{2} \theta \, \Phi \right] && + \tau_3 \, \left[ \hat \nabla^i \hat \partial_i \Phi + \delta \, \phi^i \, \hat \partial_i \Phi - \frac{\delta}{2} \left( \hat r - \frac{\delta}{2} \, \phi^i \phi_i \, \right) \, \Phi \right] \cr
&& + \sigma_3 \, \left(\hat r + a^{ij} \hat \nabla_i \phi_j \right) \, \Phi + \left( \lambda_3 + \mu_3 \, f_{ij} f^{ij} \right) \,  \Phi^{\frac{2+\delta}{\delta}} = 0
\eea
with Weyl weight $2+\delta$ and four independent parameters, and/or 
\bea
\label{z2eqns2}
\hat \partial_t^2 \Phi +  (1 + \delta) \, \theta \, \hat \partial_t \Phi && + \frac{\delta}{4} \left[ (2+\delta) \, \theta^2 + 2 \, \hat \partial_t \theta \right] \, \Phi  \cr
&& + \sigma_0 \, \left(\hat \gamma^i_j \hat \gamma^j_i -\frac{1}{2} \theta^2 \right) \, \Phi + (\lambda_0 \, + \mu_0 \, f_{ij} f^{ij}) \, \Phi^{\frac{4+\delta}{\delta}}  = 0
\eea
with Weyl weight $\delta + 4$ and three parameters.
\item For general values of $z$ we have the following two types of Carroll diffeomorphism and Weyl invariant equations of motion
\begin{enumerate}
\item 
\bea
\label{zeqns1}
&& \hat \partial_t^2 \Phi + \frac{1}{2} (z + 2 \, \delta) \, \theta \, \hat \partial_t \Phi + \frac{\delta}{4} \left[ (z+\delta) \, \theta^2 + 2 \, \hat \partial_t \theta \right] \, \Phi \cr
&& ~~~~~~ + \sigma_0 \, \left(\hat \gamma^i_j \hat \gamma^j_i -\frac{1}{2} \theta^2 \right) \, \Phi + \mu_0 \, \Phi^{1+ \frac{4}{\delta} (z-1)} f_{ij} f^{ij} +  \lambda_0 \, \Phi^{\frac{2z+\delta}{\delta}} = 0 \, ,
\eea
\item 
\bea
\label{zeqns2}
 && \hat \nabla^i \hat \partial_i \Phi + \frac{2\, \delta}{z} \, \phi^i \, \hat \partial_i \Phi - \frac{\delta}{2} \left[ \hat r - \frac{2 \, \delta}{z^2} \, \phi^i \phi_i \, \right] \, \Phi \cr
 && ~~~~~~~ + \sigma_1 \, \left(\hat r + \frac{2}{z} a^{ij} \hat \nabla_i \phi_j \right) \, \Phi + + \mu_1 \, f_{ij} f^{ij} \, \Phi^{1+ \frac{4}{\delta} (z-1)} + \lambda_1 \,  \Phi^{\frac{2+\delta}{\delta}} = 0 \, ,
 \eea
where we assume that the coefficient $\mu_0$ and $\mu_1$ are non-vanishing only if the powers of $\Phi$ multiplying them are non-negative integers.
\end{enumerate}
\end{itemize}
Now that we have derived the most general diffeomorphic and Weyl covariant equations of motion for a single real\footnote{For complex $\Phi$ one simply has to replace the potentials to be appropriate real combinations, such as $|\Phi|^{2n}$ etc.}  scalar field $\Phi$ in the background of generic Carroll geometry, we would like to turn to constructing actions for these equations next. 

We seek actions that produce each of the equations (\ref{z1eqns}, \ref{z2eqns1}, \ref{z2eqns2}, \ref{zeqns1}, \ref{zeqns2}) as their Euler-Lagrange equations for the scalar field $\Phi$. We will also restrict ourselves to the cases where (i) $\delta = \frac{z}{n_t-1}$ for some integer $n_t \ge 2$ in the {\it time-like} case and (ii) $\delta = \frac{1}{n_s-1}$ for some integer $n_s \ge 2$ for the {\it space-like} case -- as mentioned already these conditions will ensure that the potential $\Phi^N$ required will be bounded below.
\subsection{Constructing actions}
Note that our equations of motion in the absence of potentials are linear in $\Phi$ and have up to two derivatives. If we are to be able to derive them from some actions then they should be quadratic in $\Phi$ and should contain up to two derivatives. Therefore let us again start by listing all such invariants -- now counting derivatives both on the background geometric quantities $(a_{ij}, b_i, \omega)$ and on $\Phi$.
\bea
\label{listforeom}
&& \theta \, \Phi^2, ~~ \Phi \, \hat \partial_t \Phi, \cr
&& \hat \gamma^i_j \hat \gamma^j_i \, \Phi^2, ~~ \theta^2 \, \Phi^2, ~~  \theta \, \Phi \, \hat \partial_t \Phi, ~~ \hat \partial_t \theta \, \Phi^2, ~~ (\hat \partial_t \Phi)^2, ~~ \Phi \, \hat \partial_t^2 \Phi, \cr
&&  \hat r \, \Phi^2, ~~ \hat \nabla_i \phi^i \, \Phi^2, ~~  \phi^i  \phi_i \, \Phi^2, ~~ f_{ij} f^{ij} \, \Phi^2, ~~ \Phi \, \phi^i \partial_i \Phi, ~~ \hat \partial_i \Phi \hat \partial_i \Phi, ~~ \Phi \, \hat \nabla^i \hat \partial_i \, \Phi \, .
\eea
We have already listed in (\ref{weyltrans}) all the transformations required under the Carroll Weyl transformations and we simply have to find the combinations of quantities in (\ref{listforeom}) that transform homogeneously. After a straightforward analysis using the results of the previous subsection we find the following combinations up to quadratic order in $\Phi$ :
\begin{enumerate}
\item At first order in time-derivatives we have the unique combination with weight $z+2 \delta$
\bea
\label{actcan1}
\Phi \, (\hat \partial_t \Phi + \frac{\delta}{2} \theta  \, \Phi) 
\eea
along with $\Phi^{2+\frac{z}{\delta}}$ if, as before, the exponent is a positive even integer.
\item At second order in time-derivatives we find three covariant combinations with weight $2(z+\delta)$.
\bea
\label{actcan2}
\Big(\hat \partial_t \Phi + \frac{\delta}{2} \theta \, \Phi \Big)^2, ~~~~~ \Big(\hat \gamma^i_j \hat \gamma^j_i - \frac{1}{2} \theta^2 \Big) \, \Phi^2 
%
\eea
\bea
\label{actcan3}
(\hat \partial_t \theta + \frac{z}{2} \hat \gamma^i_j \hat \gamma^j_i )\, \Phi^2 + \frac{2}{\delta} \Phi \, \hat \partial_t^2 \Phi - \frac{2\delta +z}{\delta^2} (\hat \partial_t \Phi)^2 
\eea
%
%
along with $\Phi^{2+\frac{2\,z}{\delta}}$. We will refer to these as {\it time-like} combinations.
\item At second order in space-derivatives we find three combinations with Weyl weight $2 (1+ \delta)$:
\bea
\label{actcan4}
(\hat \partial_i \Phi + \frac{\delta}{z}\,  \phi_i \Phi) (\hat \partial^i \Phi + \frac{\delta}{z}\, \phi^i \Phi) 
, ~~~~~~ (\hat r + \frac{2}{z} \hat \nabla_i \phi^i ) \, \Phi^2, 
\eea
%
%
\bea
\label{actcan5}
\Phi \hat \nabla_i \hat \partial^i \Phi + 2\frac{\delta}{z} \Phi \, \phi^i \hat \partial_i \Phi + \frac{\delta}{z} (\hat \nabla_i \phi^i + \frac{\delta}{z}  \,  \phi_i  \phi^i) \, \Phi^2
\eea
along with the monomial $\Phi^{2+\frac{2}{\delta}}$. We refer to these are {\it space-like} combinations.
\item Finally we have one covariant combination with Weyl weight $4-2z + 2 \delta$ 
\bea
\label{actcan6}
f_{ij} f^{ij} \, \Phi^2 \, .
\eea
\end{enumerate} 
Now any candidate action has to be an integral over the coordinates $(t, {\bf x})$ of our Carroll manifold:
\bea
S = \int dt \, d^2{\bf x} \, {\cal L} \, .
\eea
For the action to be invariant under diffeomorphisms the Lagrangian density should transform as ${\cal L} \rightarrow {\cal L}'$ such that 
\bea
\int dt \, d^2{\bf x} ~ {\cal L} = \int dt' d^2{\bf x}' ~ {\cal L}' \, .
\eea
From (\ref{cdiff}) we have $dt' d^2{\bf x}' = J \det{J^i_j} ~ dt \, d^2 {\bf x}$. So the Lagrangian density ${\cal L}$ has to transform as ${\cal L} \rightarrow {\cal L}' = J^{-1} \det ((J^{-1})^i_j) ~ {\cal L}$ -- {\it i.e,} as a scalar density of weight $3$, equal to the dimension of the manifold -- under the relevant Carroll diffeomorphisms. The combinations we listed above in (\ref{actcan1} - \ref{actcan6}) are all scalars under Carroll diffoemorphisms, even though they have non-trivial weights under Carroll Weyl transformations. So to make them densities of suitable weights we need to multiply them by $\omega \, \sqrt{a}$ as this is the only combination of the Carroll geometry without derivatives and transforms as desired:
\bea
\omega \, \sqrt{a} \rightarrow J^{-1} \det ((J^{-1})^i_j) ~\omega \, \sqrt{a} \, ,
\eea
where $a$ is the determinant of the metric $a_{ij}$ on the base space. For instance 
\bea
\label{sampleaction}
S = \int dt \, d^2{\bf x} ~ \omega \, \sqrt{a} \, \left[\hat \partial_i \Phi \hat \partial^i \Phi + 2 \frac{\delta}{z} \Phi \,  \phi^i \hat \partial_i \Phi + \frac{\delta^2}{z^2}  \phi_i \phi^i \, \Phi^2\right],
\eea
is a good action for a Carroll diffeomorphism invariant theory. However, for it to be also a Weyl invariant theory the Lagrangian density ${\cal L}$ has to be invariant by itself (up to total-divergence terms) under (\ref{cweyl}). Let us check this for (\ref{sampleaction}): the measure transforms as $\omega \, \sqrt{a} \rightarrow B^{-2-z} \omega \, \sqrt{a}$ and the quantity in square-brackets transforms as $[\cdots ]  \rightarrow B^{2+2\delta} [\cdots]$ under the Weyl transformations (\ref{cweyl}). So demanding that the Lagrangian density ${\cal L}$ in (\ref{sampleaction}) is Weyl invariant requires 
\bea
\label{actconds1}
2+2\delta = 2+z \implies \delta = \frac{z}{2}.
\eea
This conclusion is valid for all actions constructed using the {\it space-like} combinations with Weyl weight $2+2\delta$. Similarly if we use any of the Carroll diffeomorphism invariant and Carroll Weyl covariant {\it time-like} combinations in (\ref{actcan2}) we need to fix the weight $\delta$ of the scalar such that :
\bea
\label{actconds2}
2+z = 2z+2\delta \implies \delta = 1-\frac{z}{2} \, .
\eea
Note that this is unlike the conformally coupled scalar in the background of a (pseudo) Riemann manifold where the Weyl weight of the scalar is fixed at the level of equation of motion itself (as reviewed in section \ref{sec2}), where as for the scalar in Carroll geometry it is fixed (even for a free scalar) at the level of the existence of an action. Furthermore, if we seek interactions then the potential being a monomial $\Phi^{2n}$ with positive even power puts additional constraints on the background, and hence on the weights of the fields. To see this consider a generic value of $z$ ($z$ is not necessarily equal to either 1 or 2). Then there are two types of actions where the dimensions $\delta$ of the scalar are determined as above. In each of these cases a monomial type potential term $\omega \sqrt{a} \, \Phi^{2n}$ to have the same Weyl weight as its derivative (kinetic) terms requires, as discussed earlier, to be either $\delta = \frac{z}{n_t-1}$ (in the {\it time-like} case) or $\delta = \frac{1}{n_s-1}$ (in the {\it space-like} case). Combining these with the conditions (\ref{actconds1}, \ref{actconds2}) coming from the existence of actions leads to 
\begin{itemize}
\item In the {\it time-like} case:
\bea
\delta = 1-\frac{z}{2} = \frac{z}{n_t-1} \implies z = 2 \frac{n_t-1}{n_t+1} ~~ \& ~ \delta = \frac{2}{n_t+1}, ~~ n_t \ge 2 \, .
\eea
In this case the range of $z$ is between $2/3$ (when $n_t=2$)  and $2$ (when $n_t \rightarrow \infty$). The case of $n_t =3$ gives $z=1$ with $\delta =1/2$, and this is what one obtains by taking the ultra-relativistic limit of conformally coupled scalar in 3 dimensions -- as we show in section \ref{sec4}.
\item In the {\it space-like} case:
\bea
\delta = \frac{z}{2} = \frac{1}{n_s-1} \implies z= \frac{2}{n_s-1} ~~ \& ~~ \delta = \frac{1}{n_s-1}, ~~ n_s \ge 2 \,.
\eea
This is the set of values assumed for $z$ by Duval {\it et al} \cite{Duval:2014uva}, where they denoted $k=n_s-1$ with $k\ge1$. In this case the range of $z$ is between zero (when $n_s \rightarrow \infty$) and 2 (when $n_s =2$) with the corresponding values of $\delta$ ranging between $0$ and $1$. Again the special value $z=1$ gives $\delta =1/2$.
\end{itemize}
So we conclude that the existence of (i) even and positive power monomial potentials and (ii) invariant actions implies discrete and specific rational values for both $z$ and $\delta$.
\subsection*{Actions for Equations:}

We now propose actions which produce (\ref{z1eqns} - \ref{zeqns2}) as their equations of motion. 
\begin{itemize}
\item For the {\it time-like} case with $z=2 \,(1-\delta)$ our action is
\bea
\label{tlikeact}
S_t = \int dt \, d^2{\bf x} ~~  \omega \sqrt{a} \left[\alpha_1 \, \Big(\hat \partial_t \Phi + \frac{\delta}{2} \theta \, \Phi \Big)^2 + \beta_1 \, \Big(\hat \gamma^i_j \hat \gamma^j_i - \frac{1}{2} \theta^2 \Big) \, \Phi^2  + \lambda_1 \, \Phi^{2+\frac{2\,z}{\delta}}\right], 
\eea
which gives the equation of motion for $\Phi$ to be:
\bea
&& -2 \, \alpha_1 \left[ \hat \partial_t^2 \Phi + \theta \, \hat \partial_t \Phi -\frac{1}{4} \delta \, (\delta-2) \, \theta^2 \, \Phi  + \frac{\delta}{2} \hat \partial_t \theta \, \Phi \right] \cr
&& + 2 \beta_1 \, \Big(\hat \gamma^i_j \hat \gamma^j_i - \frac{1}{2} \theta^2 \Big) \, \Phi +2\, \lambda_1 \, \left(\frac{2}{\delta}-1\right) \, \Phi^{\frac{4}{\delta}-3} =0 
\eea
\item For the {\it space-like} case with $z=2\delta$ our action is
\bea
\label{slikeact}
S_{s} = \int dt \, d^2{\bf x} ~ \omega \, \sqrt{a} \, \left[\alpha_2 \, \Big(\hat \partial_i \Phi + \frac{\delta}{z}\, \phi_i \Phi \Big) \Big(\hat \partial^i \Phi + \frac{\delta}{z}\, \phi^i \Phi \Big) 
+ \beta_2 \, \Big(\hat r + \frac{2}{z} \hat \nabla_i  \phi^i \Big) \, \Phi^2 +\lambda_2 \, \Phi^{2+\frac{2}{\delta}}\right], \cr
\eea
which gives rise to
\bea
&& -2 \, \alpha_2 \left[ \hat \square \Phi +  \phi^i \hat \nabla_i \Phi + \frac{1}{4}  \phi^i \phi_i \Phi + \frac{1}{2} \hat \nabla_i  \phi^i \, \Phi \right] \cr
&& + 2\beta_2 \, \Big(\hat r + \frac{2}{z} \hat \nabla_i  \phi^i \Big) \, \Phi + 2 \lambda_2 \Big(1+\frac{1}{\delta} \Big) \, \Phi^{1+\frac{2}{\delta}} =0 \, .
\eea

\end{itemize}
It is easy to see that these equations are the same as (\ref{zeqns1} - \ref{zeqns2}) with appropriate identifications of the coefficients.

For the special case $z=1$ we can consider linear combinations of (\ref{tlikeact}), (\ref{slikeact}) with (\ref{actcan6}), which can be seen to generate equations of the form (\ref{z1eqns}). The other special value of $z$, namely $z=2$ needs some more attention. In this case we either have to consider a complex $\Phi$ or two real fields $\Phi_1$ and $\Phi_2$. 


%
%

\section{A Carroll CFT from a conformally coupled scalar}
\label{sec4}

Now we turn to show that some special cases of the equations and their corresponding actions we derived in the last section arise in the ultra-relativistic limit of the conformally coupled scalar reviewed in section \ref{sec2}.

Our starting point is the diffeomorphic and Weyl invariant scalar field theory in $d+1$ dimensional spacetime. The background metric is taken in the so-called Randers-Papapetrou form (see for instance \cite{Ciambelli:2018xat}),
%
%
with the line element:
\bea
\label{rpmetric}
ds^2 =  g_{\mu\nu} dx^\mu dx^\nu = - c^2 \omega^2 (dt - \omega^{-1} b_i \, dx^i)^2 + a_{ij} \, dx^i \, dx^j
\eea
where $x^\mu = (t, {\bf x})$. This geometry in the $c \rightarrow 0$ limit is expected to produce a Carroll geometry. Moreover the subset of all diffeomorphisms that leave this metric form-invariant are precisely the Carroll diffeomorphisms (\ref{cdiff}) and the quantities $\{a_{ij}, b_i, \omega\}$ transform under (\ref{cdiff}) as in (\ref{cdifftrans}). 

It can be seen that the Ricci scalar of this geometry is
\bea
R = c^{-2} \left( \theta^2 + \hat \gamma^i_j \hat \gamma^j_i + 2 \, \hat \partial_t \theta \right) + \left(\hat r - 2 \hat \nabla_i \phi^i - 2 \,  \phi_i \phi^i \right) + \frac{c^2}{4} f_{ij} f^{ij}  \, .
\eea
Consider the Weyl invariant Klein-Gordon scalar field equation on a general background in three dimensions:
\begin{equation}
\label{kgagain}
g^{ab}\nabla_a \nabla_b\, \Phi-\frac{1}{8}R\,\Phi=0
\end{equation}
Using Randers-Papapetrou metric ansatz (\ref{rpmetric}) one can reduce this equation to a polynomial in $c$. This equation admits an expansion in terms of the combinations in (\ref{z1eqns} - \ref{zeqns2}) with $(z, \delta) = (1, \frac{1}{2})$. In particular:
\bea
\label{kginrp}
\hat \square \Phi - \frac{1}{8} \, R \, \Phi &&\cr
= &-&\frac{1}{c^2} \left[ \hat \partial_t^2 \Phi + \theta \, \hat \partial_t \Phi + \frac{1}{16} \left( 3 \, \theta^2 + 4 \, \hat \partial_t \theta \right) \, \Phi \right] - \frac{1}{8 \, c^2} \left( \hat \gamma^i_j \hat \gamma^j_i - \frac{1}{2} \theta^2 \right) \cr
&+& \left[ \hat \square \Phi +  \phi^i \hat \nabla_i \Phi - \frac{1}{4} \left( \hat r -  \phi^i \phi_i \right) \Phi \right] + \frac{1}{8} \left( \hat r + 2 \, \hat \nabla_i \phi^i \right) \, \Phi \cr
&-& \frac{c^2}{32} f_{ij} f^{ij} \, \Phi \, .
\eea
Notice that if we had set any of these terms at any given order in powers of $c$ to zero, it would have given us a Carroll diffeomorphic and Weyl covariant equation. We do not have to just restrict to the lowest order term (the leading term in the ultra-relativistic limit -- as was done in the previous works \cite{Duval:2014uoa, Duval:2014uva, Duval:2014lpa, Ciambelli:2019lap, Ciambelli:2018wre, Ciambelli:2018xat, Bagchi:2019clu, Bagchi:2019xfx, Bagchi:2016bcd, Ciambelli:2018ojf}). However, if we took the linear combination exactly as in (\ref{kginrp}) we can re-package the equation into a fully diffeomorphic and Weyl covariant equation (\ref{kgagain}) in the pseudo-Riemannian space with metric (\ref{rpmetric}).


Let us now turn to the action in this special case. For this we start with the Lagrangian density of the conformally coupled scalar to the background Randers-Papapetrou metric (\ref{rpmetric}) and again expand it in powers of $c$. This gives us:
\bea
S =&& \frac{1}{c} \int dt \, d^2{\bf x} ~ \omega \sqrt{a} \left[ (\hat \partial_t \Phi)^2 - \frac{1}{8} \Big(\theta^2 + \hat \gamma^i_j \hat \gamma^j_i + 2 \, \hat \partial_t \theta \Big) \, \Phi^2 \right] \cr
&& -c \int dt \, d^2{\bf x} ~ \omega \sqrt{a} \left[ \hat \nabla_i \Phi \, \hat \nabla^i \Phi + \frac{1}{8} ( \hat r - 2 \, \hat \nabla_i \phi^i - 2 \,  \phi_i  \phi^i) \, \Phi^2 \right] \cr
&& - \frac{c^3}{32} \int dt \, d^2{\bf x} ~ \omega \sqrt{a} \, f_{ij} f^{ij} \, \Phi^2 \, .
\eea
If we put $\alpha_1 = \frac{1}{c}$ and $\beta_1 = -\frac{1}{8c}$ (with $\lambda_1 =0$) and use 
\bea
\partial_t (\sqrt{a} \, \theta \, \Phi^2) = \omega \, \sqrt{a} \, \left[ 2 \, \theta \, \Phi \, \hat \partial_t \Phi + (\theta^2 + \hat \partial_t \theta) \, \Phi^2 \right]
\eea
we see that the ${\cal O}(1/c)$ term of this action is a special case of (\ref{tlikeact}) with $(z, \delta) = (1, \frac{1}{2})$ up to a total derivative. Similarly noticing that 
\bea
\partial_i \left[ \omega \sqrt{a} \,  \phi^i \, \Phi^2 \right] + \partial_t \left[ \sqrt{a} \, b_i  \phi^i \, \Phi^2 \right] =\omega \, \sqrt{a} \, \left[ 2 \,\Phi  \phi^i \hat \partial_i \Phi +  ( \phi^i  \phi_i + \hat \nabla_i  \phi^i) \, \Phi^2 \right]
\eea
and setting $\alpha_2=1$, $\beta_2 = \frac{1}{8}$ we can see that the ${\cal O}(c)$ term in this action is again a special case of (\ref{slikeact}). Finally the order $c^3$ term can be added for free again in this case of $z=1$ and $\delta=1/2$ as before.

\section{Gauge fixing and residual symmetries}
 \label{sec5}
So far we have been using both Carroll diffeomorphisms and Carroll Weyl symmetries to constrain our theories. However, our main interest is in getting down to Carrollian CFTs. For this we need to fix the background spacetime geometry -- as much as possible -- using the local symmetries. In the standard construction of CFTs in the background of a (pseudo) Riemannian manifold one takes a fixed background metric (or a representative in its conformal class). Then the local symmetries (in $(d+1)$-dimensional case) can be used to fix $d+2$ components of the metric completely. This allows one to gauge fix any metric to a specific member of its conformal class. Furthermore if one wants to get a conformal field theory with symmetry algebra $so(2, d+1)$ then the background should be conformally flat. This condition requires that the metric $g_{\mu\nu}$ should have vanishing Weyl tensor: $W_{\mu\nu\sigma\lambda} =0$ (or the Cotton tensor $C_{\mu\nu} =0$ in $d=2$ case).

Turning now to the Carroll case we again have in a generic geometry, specified by $(a_{ij}, b_i, \omega)$, as many as $\frac{1}{2} (d+1) (d+2)$ components which are all arbitrary functions of $(t, {\bf x})$. But the local symmetries have two functions ($t'(t,{\bf x})$, $B(t, {\bf x})$) of both space and time and $d$ functions ($x'^i({\bf x})$) of space alone. 

We are looking for 3-dimensional Carrollian CFTs with symmetry algebra $\mathfrak{cca}_3^{(z)}$ that contains the conformal algebra of the two dimensional base space $so(1,3)$ -- so we would like to be able to completely gauge fix $a_{ij}$ to some fixed time-independent one. One possibility is that we restrict to Carroll geometries where $a_{ij}$ is of the form:
\bea
\label{shearcondition}
a_{ij} (t, {\bf x}) = e^{\chi(t, {\bf x})} a^{(0)}_{ij} ({\bf x}), ~~~ \det a^{(0)}_{ij}({\bf x}) =1
\eea
In other words the metric $a_{ij}$ is conformally time-independent. Then we will have two components in $a^{(0)}_{ij}({\bf x})$ which can be gauge fixed completely using the spatial diffeomorphisms alone. Then we can use the temporal diffeomorphism and the Weyl symmetry to gauge fix $\chi(t, {\bf x}) =0$ and $\omega(t, {\bf x}) =1$. It turns out that this will be sufficient to ensure that the residual symmetry algebra to be $\mathfrak{cca}_3^{(z)}$. Also the covariant condition (analogous to vanishing Weyl or Cotton tensors in the pseudo-Riemannian case) is $\hat \gamma_{ij} - \frac{1}{2} \theta \, a_{ij} =0$. 

%
Some examples of Carroll spacetimes include: Flat Carroll spacetime \cite{Duval:2014uoa, Duval:2014uva, Duval:2014lpa, Ciambelli:2019lap} given by
\bea
a_{ij} = \delta_{ij}, ~~ b_i = b_i^{(0)}, ~~ \omega =1 
\eea
where $b_i^{(0)}$ are constants. For this we have $\theta = \phi_i = \gamma^i_j = f_{ij} = \hat \gamma^i_{jk} =0$. The {\it time-like} action (\ref{tlikeact}) is what people obtained by ultra-relativistic limit of the Klein-Gordon scalar in 3-dimensional Minkowski spacetime $M_{2+1}$. The {\it space-like} action (\ref{slikeact}) becomes, for $z=2\delta$
\bea
\int dt \, d^2{\bf x} ~ \left[\hat \partial_i \Phi \hat \partial_i \Phi + \lambda \, \Phi^{2+\frac{2}{\delta}}\right] \, .
\eea
More general Carroll manifolds that include some of the interesting Carroll spacetimes, such as, null infinities, black hole horizons etc, and their conformal symmetries are discussed in detail in \cite{Ciambelli:2019lap}. To study the symmetries of the gauge fixed actions one takes the vector field that generates Carroll diffeomorphism to be of the form:
\bea
\xi &=& f(t, {\bf x}) \, \hat \partial_t + \xi^i ({\bf x}) \, \hat \partial_i \cr
&=& \omega^{-1} \, (f + \xi^i \, b_i)  \, \partial_t + \xi^i \partial_i \, .
\eea
Under the infinitesimal coordinate transformation 
\bea
t' = t + \omega^{-1} \, (f + \xi^i \, b_i) + \cdots, ~~ x'^i = x^i + \xi^i  + \cdots
\eea
the background data $(a_{ij}, b_i, \omega)$ and of matter field $\Phi$ transform as:
\bea
\label{infdiff}
\delta_\xi a_{ij} &=& -2 f\, \hat \gamma_{ij} - (\hat \nabla_i \xi_j + \hat \nabla_j \xi_i) \, , \cr
\omega^{-1} \, \delta_\xi \omega &=& - \hat \partial_t f - \phi_i \, \xi^i \, , \cr
\delta_\xi b_i &=& -b_i \, (\phi_j \xi^j + \hat \partial_t f) + f_{ij} \, \xi^j  + (\hat \partial_i - \phi_i ) \,f \, , \cr
\delta_\xi \Phi &=& - (f \, \hat \partial_t \Phi + \xi^i \hat \partial_i \Phi) \, .
\eea
Under the infinitesimal Weyl transformations with $B = e^\sigma$, we have
\bea
\label{infweyl}
\delta_\sigma a_{ij} = -2 \, \sigma \, a_{ij}, ~~ \delta_\sigma \omega = -z \, \sigma \, \omega, \cr
\delta_\sigma b_i = -z \, \sigma \, b_i, ~~ \delta_\sigma \Phi = \delta \, \sigma \, \Phi \, .
\eea
One first demands that the metric on the base $a_{ij}$ is invariant under the combined action (\ref{infdiff}, \ref{infweyl}):
\bea
(\delta_\xi + \delta_\sigma) \, a_{ij} =0
\eea
and this leads to
\bea
2\, f \, \Big[ \hat \gamma_{ij} - \frac{1}{2} \theta \, a_{ij} \Big] + \hat \nabla_i \xi_j + \hat \nabla_j \xi_i - \hat \nabla_k \xi^k \, a_{ij} =0, ~~~ \sigma = -\frac{1}{2} ( f \, \theta + \hat \nabla_i \xi^i) \, .
\eea
Now following \cite{Ciambelli:2019lap} we will also choose to impose that the traceless symmetric tensor $\zeta_{ij} = \hat \gamma_{ij} - \frac{1}{2} \theta \, a_{ij}$ (referred to as the Carroll shear) vanishes. This condition can be solved for and it implies (\ref{shearcondition}) and hence $a_{ij}$ is conformally time independent. Next we choose to impose $(\delta_\xi + \delta_\sigma) \omega =0$ and this leads to 
\bea
(\hat \partial_t  - \frac{z}{2}\theta) \, f - \frac{z}{2} (\hat \nabla_i - \frac{2}{z} \phi_i) \, \xi^i &=& 0 \, , \cr \cr
{\rm along ~ with} ~~~~~~~ \hat \nabla_i \xi_j + \hat \nabla_j \xi_i - \hat \nabla_k \xi^k \, a_{ij} &=& 0 \, .
\eea
Finally noticing that these equations are Carroll Weyl invariant one chooses $a_{ij}$ to be completely time independent fixed metric. This implies that $\theta =0$ and the Carroll Levi-Civita connection $\hat \gamma^i_{jk}$ reduces to the Christoffel connection for (now time independent) $a_{ij}$. One also chooses a fixed background value for $\omega$ (say, $\omega=1$) without loss of generality. Then the residual symmetries have to satisfy 
\bea
\nabla_i \xi_j + \nabla_j \xi_i - \nabla_k \xi^k \, a_{ij} &=& 0 \, ,\cr
\partial_t \, f - \frac{z}{2} (\nabla_i - \frac{2}{z} \phi_i) \, \xi^i &=& 0 \, .
\eea
One can integrate these equations completely and the result is given in terms of $\{T({\bf x}), Y^i({\bf x}\}$ where $\xi^i =Y^i({\bf x})$ are the conformal Killing vectors on $a_{ij}$ and $T({\bf x})$ is arbitrary:
\bea
\label{fsoln}
f(t,{\bf x}) &=& T({\bf x}) + \frac{z}{2}\int^t dt' \left[ \nabla_i Y^i({\bf x})- \frac{2}{z} \phi_i \, Y^i({\bf x})  \right] \cr
&=& T({\bf x}) + \frac{z}{2} t \, \nabla_i Y^i({\bf x}) - b_i (t, {\bf x}) Y^i({\bf x})
\eea
where we have set $\omega(t, {\bf x}) =1$ and $\phi_i = \partial_t b_i$.  So the Carrollian conformal Killing vector is 
\bea
\label{finalcckv}
\xi &=& (f + \xi^i \, b_i)  \, \partial_t + \xi^i \partial_i \cr
&=& \left[T({\bf x}) + \frac{z}{2} t \, \nabla_i Y^i({\bf x}) \right] \, \partial_t + Y^i({\bf x}) \, \partial_i \, .
\eea
It is argued in \cite{Ciambelli:2019lap} that the algebra of the Carrollian CKV do not depend on the choice of $\omega$ either. The fact that the $b_i$ dependence canceled out in the final answer (\ref{finalcckv}) is also general enough that we do not need to fix $b_i$ to define the residual symmetries. Therefore the residual fields $b_i(t, {\bf x})$ along with $\Phi$ transform under the Carrollian conformal transformations as:
\bea
\label{phibtrans}
\delta b_i &=& -b_i \, (\phi_j \xi^j + \hat \partial_t f) + f_{ij} \, \xi^j  + (\hat \partial_i - \phi_i ) \,f -z \, \sigma \, b_i \, ,\cr
\delta \Phi &=& - (f \, \hat \partial_t \Phi + \xi^i \hat \partial_i \Phi) + \delta \, \sigma \, \Phi \, ,
\eea
where we have to use $a_{ij}$ to be fixed and time independent (for instance that of a round $S^2$), $\omega =1$ (and hence $\phi_i = \partial_t b_i$), $\xi^i = Y^i({\bf x})$ and $f(t, {\bf x})$ as in (\ref{fsoln}). These form the symmetries of our theories by construction when we gauge fix as above. This means that our $\mathfrak{cca}_3^{(z)}$ symmetric scalar field theories, after gauge fixing the Carroll diffeomorphisms and Weyl symmetries would have the matter field $\Phi(t,{\bf x})$ along with the  geometric fields $b_i(t, {\bf x})$ dynamical. Then the transformations (\ref{phibtrans}) are their symmetries, and can be used to study the conserved currents and charges {\it etc.} of our theories both at classical level and beyond. 

There may be other interesting possibilities to gauge fix the local symmetries with apparently different residual symmetry algebras - we discuss one such gauge now. Let us consider restricting the background Carroll geometric data $(a_{ij}, b_i, \omega)$ by imposing $\phi_i =0$. This condition can be solved as follows:
\bea
\frac{1}{\omega}(\partial_i \omega + \partial_t b_i) =0 \implies b_i (t, {\bf x}) = b_i^{(0)} ({\bf x}) - \partial_i \int^t dt' \, \omega (t', {\bf x})
\eea
If we further gauge fix $\omega =1$ as in the previous gauge choice we conclude that $\phi_i =0$ implies that $b_i$ is independent of time. So we can use the spatial Carroll diffeomorphisms to gauge fix $b_i = 0$.\footnote{A weaker condition is to fix $b_i$ to be such that $f_{ij}=0$.} Finally we can use the Weyl symmetry to fix the determinant of $a_{ij}$, say $a =1$, which in turn will imply $\theta =0$. So in this gauge we have $\phi_i = \theta = b_i = f_{ij} =0$. Using $(\delta_\xi + \delta_\sigma) \, b_i =0$ leads to $\partial_i f=0$. From $(\delta_\xi + \delta_\sigma) \, \omega =0$ we find $\sigma = - \frac{1}{z}\partial_t f(t)$. This gauge leaves $a_{ij}(t, {\bf x})$ to fluctuate subject to the condition $a =1$ along with $\Phi(t, {\bf x})$. The condition $a=1$ also implies $\sigma =- \frac{1}{2} \nabla_i \xi^i$. Since $\sigma \sim \partial_t f(t)$ the only consistent choices are $f(t) = \alpha + \beta \, t$ and $\nabla_i \xi^i =\frac{2}{z}\beta$, for constant $\alpha$ and $\beta$. One may further choose to set $\beta =0$ (though this is not necessary in all the cases) as in \cite{Compere:2019bua}. Then one has $\xi = \alpha \, \partial_t + \xi^i \partial_i$ with constant $\alpha$ and $\nabla_i \xi^i =0$, giving rise to an algebra isomorphic to ${\mathbb R} \times {\cal A}$ where ${\cal A}$ is the algebra of volume-preserving (smooth) diffeomorphisms of $a_{ij}$ (say ${\mathbb R}^2$ or round $S^2$). In the $z=1$ case such an algebra has appeared recently \cite{Compere:2019bua} in a different context. 

We anticipate that it should be possible to gauge fix the background such that one obtains the extension of $\mathfrak{bms}_4$ considered in \cite{Campiglia:2014yka}. We will leave further studies of these aspects to a future publication.
 
\section{Conclusion}
\label{sec6}

We have derived equations for a scalar field coupled to a generic 3-dimensional Carroll geometry that are Carroll diffeomorphic and Weyl covariant. These exist for any value of the dynamical exponent $z$ and conformal dimension $\delta$ of the scalar. We have also shown how to construct corresponding actions for such theories and analysed the consequences for $z$ and $\delta$. After an appropriate gauge fixing we found that the field theories we sought, with Carrollian conformal symmetries have in addition to $\Phi(t, {\bf x})$ two more dynamical fields coming from $b_i(t, {\bf x})$ in their field content. The {\it space-like} theories resemble more (deformations) of standard Euclidean CFTs with additional features than the {\it time-like} theories. 

Even though we have demonstrated our methods most explicitly for 3-dimensional Carroll spacetimes the generalisation to arbitrary (higher) dimensions is straightforward -- see the Appendix \ref{appendix} for some technical details of this exercise. Also one can include Carroll like Maxwell, Yang-Mills fields and other matter fields following similar methods -- these details will be presented in \cite{GS2}. 

We have concentrated on classical theories in this paper. It will be interesting to explore the quantum aspects of our theories -- including renormalisation, anomalies etc. We need to compute the Noether charges for the symmetries and show that they form $\mathfrak{cca}_3^{(z)}$. In particular we will need to construct the soft charges -- which may lead to Ward identities of celestial amplitudes in connection with the soft graviton theorems (see \cite{Strominger:2017zoo} for a review). 

There are some features expected of holographic duals of flat space gravity \cite{Banerjee:2019aoy, Banerjee:2019tam}. It will be interesting to see if these are borne out in theories of the type we constructed here. In this connection, we anticipate that the existence of additional fields, either $b_i(t, {\bf x})$ (or $a_{ij}(t, {\bf x})$ with $a=1$) may play a useful role. One also expects such requirements to impose further cuts on the spaces of classical theories we find, along with any other possible quantum consistency conditions.

In this work we concentrated on Carrollian theories. However, the same techniques can be used for constructing the Galilean theories as well. There is a curious duality between Carrollian and Galilean field theories \cite{Duval:2014uoa}. It would be interesting to check if such duality exists between our theories and the corresponding Galilean ones. We hope to present results on this in the near future. 

Eventually one would like to construct fully consistent (perhaps supersymmetric) Carrollian CFTs which can potentially be useful to describe flat space gravity/string theories holographically. We hope that our methods would lead to more avenues to explore than the method of taking ultra-relativistic limits of known theories.

\section*{Acknowledgements}
We thank participants of Chennai Strings Meeting 2019 (Nov 21-23) at IMSc for their feedback on a talk by NVS on this topic. 

\appendix
\section{Extension to general dimensions -- some details}
\label{appendix}
{\small
In this appendix we provide some technical details of the extension of the $d=2$ computations of this paper to general $d$. Let us start by providing the transformation properties of various Carroll diffeomorphic quantities under Carroll Weyl transformations - generalising those of (\ref{wtrans1}).
\begin{align*}
\theta&\longrightarrow B^{z-1}\bigg[B\,\theta-d\,\hat{\partial}_t B\bigg], ~~
\theta^2 \longrightarrow B^{2z-2}\bigg[B^2\theta -2B\theta\,d\,\hat{\partial}_t B+d^2(\hat{\partial}_tB)^2\bigg]\\
\hat{\gamma}^i_j\hat{\gamma}^j_i&\longrightarrow B^{2z-2}\bigg[B^2\hat{\gamma}^i_j\hat{\gamma}^j_i-2B\,\theta\hat{\partial}_tB+d\,(\hat{\partial}_tB)^2\bigg] \\
\hat{r}&\longrightarrow B^2\,\hat{r}-d(d-1)a^{ij}\hat{\nabla}_iB\,\hat{\nabla}_jB+2(d-1)a^{ij}B\,\hat{\nabla}_i\hat{\nabla}_jB \\
a^{ij}\hat{\nabla}_i\phi_j&\longrightarrow B^2\,a^{ij}\hat{\nabla}_i\phi_j+z(d-1)a^{ij}\hat{\nabla}_iB\,\hat{\nabla}_jB-z\,B\,a^{ij}\hat{\nabla}_i\hat{\nabla}_jB-B(d-2)a^{ij}\phi_i\hat{\nabla}_jB\\
a^{ij}\phi_i\phi_j&\longrightarrow B^2a^{ij}\phi_i\phi_j-2zBa^{ij}\phi_i\hat{\nabla}_jB+z^2a^{ij}\hat{\nabla}_iB\,\hat{\nabla}_jB
\end{align*}
The combinations that transform homogeneously are
\bea
&&\hat{r}+\frac{2}{z}(d-1)a^{ij}\hat{\nabla}_i\phi_j-\frac{(d-2)(d-1)}{z^2}a^{ij}\phi_i\phi_j \cr 
&& ~~~~~~~~~~ \longrightarrow B^2\bigg(\hat{r}+\frac{2}{z}(d-1)a^{ij}\hat{\nabla}_i\phi_j-\frac{(d-2)(d-1)}{z^2}a^{ij}\phi_i\phi_j\bigg) \cr
&&\hat{\gamma}^i_j\hat{\gamma}^j_i-\frac{\theta^2}{d} \longrightarrow B^{2z}\bigg[\hat{\gamma}^i_j\hat{\gamma}^j_i-\frac{\theta^2}{d}\bigg]
\eea
Taking that the scalar $\Phi$ transforms as $\phi\rightarrow B^{\delta}\Phi$ under Weyl transformations various expression transforms as follows
\begin{align*}
a^{ij}\phi_{i}\hat{\partial}_j\Phi&\longrightarrow B^{\delta}\big[ B^2\, a^{ij}\phi_j\hat{\partial}_i\Phi+\delta B\,a^{ij}\phi_i\hat{\partial}_j B\, \Phi-\delta z a^{ij}\hat{\nabla}_iB\hat{\nabla}_jB\,\Phi-zBa^{ij}\hat{\partial}_i B\hat{\partial}_j\Phi\big] \\
\hat{\nabla}_i\hat{\partial}^i\Phi&\longrightarrow B^{\delta}\big[B^2\hat{\nabla}_i\hat{\partial}^i\Phi+(\delta^2-d\delta+\delta)a^{ij}\hat{\nabla}_i B\hat{\nabla}^iB\,\Phi+\delta B\hat{\nabla}_i\hat{\nabla}^iB\, \Phi+ (2+2\delta-d)B\,\hat{\partial}_i B\, \hat{\partial}^i\Phi\big]
\end{align*}
The simplest Weyl covariant object at first order derivative is 
\bea
\hat{\partial}_t\Phi+\frac{\delta}{d}\theta\, \Phi \longrightarrow B^{z+\delta}(\hat{\partial}_t\Phi+\frac{\delta}{d}\theta\, \Phi)
\eea
At second order we find:
\bea
&&\hat{\partial}^2_t \Phi+\frac{1}{d}(2\delta+z)\,\theta\, \hat{\partial}_t\Phi+\frac{\delta}{d}\bigg[\frac{1}{d}(z+\delta)\,\theta^2+\hat{\partial}_t\theta\bigg] \,\Phi \cr
&& ~~~~~~~~ \longrightarrow B^{2z+\delta}\bigg(\hat{\partial}^2_t+\frac{1}{d}(2\delta+z)\,\theta\, \hat{\partial}_t\Phi+\frac{\delta}{d}\bigg[\frac{1}{d}(z+\delta)\,\theta^2+\hat{\partial}_t\theta \,\Phi\bigg]\bigg)\\
&& \hat{\nabla}^i\hat{\partial}_i\Phi+\frac{(2-d+2\delta)}{z}\phi^i\hat{\partial}_i\Phi-\frac{\delta}{2}\bigg[\frac{1}{d-1}\hat{r}-\frac{1}{z^2}(2-d+2\delta)\phi^i\phi_i\bigg]\Phi \cr 
&& ~~~~~~~~ \longrightarrow B^{2+\delta}\bigg(\hat{\nabla}^i\hat{\partial}_i\Phi+\frac{(2-d+2\delta)}{z}\phi^i\hat{\partial}_i\Phi-\frac{\delta}{2}\bigg[\frac{1}{d-1}\hat{r}-\frac{1}{z^2}(2-d+2\delta)\phi^i\phi_i\bigg]\Phi\bigg)
\eea
For $z=1$ the covariant equation can be written as,
\bea
&& \kappa_0\bigg[\hat{\partial}^2_t\Phi+\frac{1}{d}(2\delta+z)\,\theta\, \hat{\partial}_t\Phi+\frac{\delta}{d}\bigg[\frac{1}{d}(z+\delta)\,\theta^2+\hat{\partial}_t\theta\bigg] \,\Phi\bigg]\cr
&& +\kappa_1\bigg[\hat{\nabla}^i\hat{\partial}_i\Phi+\frac{(2-d+2\delta)}{z}\phi^i\hat{\partial}_i\Phi-\frac{\delta}{2}\bigg[\frac{1}{d-1}\hat{r}-\frac{1}{z^2}(2-d+2\delta)\phi^i\phi_i\bigg]\Phi\bigg]\cr
&& + \bigg[\sigma_0\bigg(\hat{\gamma}^i_j\hat{\gamma}^j_i-\frac{\theta^2}{d}\bigg)+\sigma_1\bigg(\hat{r}+\frac{2}{z}(d-1)a^{ij}\hat{\nabla}_i\phi_j-\frac{(d-2)(d-1)}{z^2}a^{ij}\phi_i\phi_j\bigg)+\sigma_2 f_{ij}f^{ij}\bigg]\Phi+\lambda\Phi^{\frac{2+\delta}{\delta}} =0 \nonumber \\
\eea
The general $d$ expressions of invariants in (\ref{actcan1}-\ref{actcan6}) that can be used to construct actions are 
\begin{align}
\Phi(\hat{\partial}_t\Phi+\frac{\delta}{d}\,\theta\, \Phi)\\
\big(\hat{\partial}_t\theta+\frac{z}{2}\hat{\gamma}^i_j\hat{\gamma}^j_i\big)\Phi^2+\frac{d}{\delta}\Phi\hat{\partial}^2_t\Phi-\frac{d(z+2\delta)}{2\delta^2}(\hat{\partial}_t\Phi)^2\\
(\hat{\partial}_t\Phi)^2+\delta\,\theta\Phi\hat{\partial}_t\Phi+\frac{{\delta}^2}{d}{\gamma}^i_j\hat{\gamma}^j_i\Phi^2
\end{align}
At second order in space derivatives there are three more combinations with Weyl weight $2+2\delta$:
\begin{align}
\bigg[\hat{r}+\frac{2}{z}(d-1)a^{ij}\hat{\nabla}_i\phi_j-\frac{(d-2)(d-1)}{z^2}a^{ij}\phi_i\phi_j\bigg]\Phi^2\\
\hat{\partial}_i\Phi\hat{\partial}^i\Phi+2\frac{\delta}{z}\Phi\phi^i\hat{\partial}_i\Phi+\frac{{\delta}^2}{z^2}\phi^i\phi_i\Phi^2 \\
\Phi\hat{\nabla}^i\hat{\partial}_i\Phi-\frac{d-2(1+\delta)}{z}\Phi\phi^i\hat{\partial}_i\Phi+\frac{\delta}{z}\hat{\nabla}^i\phi_j\Phi^2+\frac{\delta}{z^2}(2-d+\delta)\phi^i\phi_i\Phi^2
\end{align}
These can be used to construct Carroll diffeomorphic and Weyl invariant actions for general $d$ straightforwardly. 
}

\bibliographystyle{utphys}
\providecommand{\href}[2]{#2}\begingroup\raggedright\endgroup
\end{document}